\documentclass[twocolumn,showpacs,amsmath,amssymb]{revtex4}
\usepackage{amsmath}
\usepackage{graphicx}
\usepackage{amsfonts}
\usepackage{dcolumn}
\usepackage{bm}

\begin{document}

\title{Noise-induced vortex reversal of self-propelled particles}

\author{Hanshuang Chen$^{1,2}$}

\author{Zhonghuai Hou$^{1}$}\email{hzhlj@ustc.edu.cn}

\affiliation{$^{1}$Hefei National Laboratory for Physical Sciences
at Microscales \& Department of Chemical Physics, University of
 Science and Technology of China, Hefei, 230026, People¡¯s Republic of China \\ $^{2}$School of Physics and Material Science, Anhui
University, Hefei, 230039, People¡¯s Republic of China }

\date{\today}

\begin{abstract}
We report an interesting phenomenon of noise-induced vortex reversal
in a two-dimensional system of self-propelled particles (SPP) with
soft-core interactions. With the aid of forward flux sampling, we
analyze the configurations along the reversal pathway and thus
identify the mechanism of vortex reversal. We find that
statistically the reversal exhibits a hierarchical process: those
particles at the periphery first change their motion directions, and
then more inner layers of particles reverse later on. Furthermore,
we calculate the dependence of the average reversal rate on noise
intensity $D$ and the number $N$ of SPP. We find that the rate
decreases exponentially with the reciprocal of $D$. Interestingly,
the rate varies nonmonotonically with $N$ and a minimal rate exists
for an intermediate value of $N$.

\end{abstract}
 \maketitle

In recent years, the collective dynamics of self-propelled particles
(SPP) has been a subject of intense research due to the potential
implications in biology, physics, and engineering (see
\cite{arXiv1010.5017} for a recent review). Examples of SPP are
abundant including traffic flow \cite{RevModPhys.73.1067}, birds
flocks \cite{PNAS107.11865,NJP12.093019}, insects swarms
\cite{Science312.1402,PhysRevLett.102.010602}, bacteria colonies
\cite{PhysRevLett.102.048104,PNAS107.13626}, and active granular
media \cite{Science317.105,PhysRevLett.105.098001}, just to list a
few. Especially, inspired by the seminal work of Vicsek \emph{et
al.} \cite{PhysRevLett.75.1226}, many theoretical and experimental
studies reported that various systems of SPP can exhibit a wealth of
emergent nonequilibrium patterns like swarming, clustering and
vortex
\cite{PhysRevLett.75.4326,PhysRevLett.92.025702,PhysRevLett.98.095702,
PhysRevLett.96.180602,PhysRevE.63.017101,PhysRevLett.96.104302,PhysRevLett.106.128101}.

A fascinating phenomenon about SPP is that they can abruptly change
their collective motion pattern, which may be induced by either
intrinsic stochasticity such as error in communication among SPP or
a response to an external influence such as a predator. For example,
a recent experiment showed that marching locusts can suddenly switch
their direction without any change in the external environment
\cite{Science312.1402}. Later, the experimental results are further
explained theoretically by a mathematical modeling, highlighting the
nontrivial role of randomness or noise on this transition
\cite{PNAS106.5464}. Also, noise-induced transitions between
translational motion and rotational motion of SPP have been observed
\cite{PhysRevE.60.4571,PhysRevE.71.051904,PNAS104.5931,PhysRevE.78.031927}.
However, the research on this topic is still in its infancy and
deserves more investigations. In particular, investigation of the
mechanisms about these transitions is lacking at present. As we
know, identifying the underlying mechanisms is key to understanding
and controlling the collective motion of SPP.

In this paper, we report an interesting phenomenon of noise-induced
vortex reversal between two different rotational directions in a
two-dimensional model of SPP interacting via Morse potential
\cite{PhysRevLett.96.104302}. Vortex pattern has been commonly
observed in nature, such as fish, ants
\cite{BiolBull202.296,PhilTransRSocB361.5}, \emph{Bacillus subtilis}
\cite{PhysRevE.54.1791}, and \emph{Dictyostelium} cells
\cite{PhysRevLett.83.1247}. By virtue of a recently developed
simulation method of rare event, forward flux sampling (FFS)
\cite{PhysRevLett.94.018104}, we analyze the intermediate
configurations along the reversal path and compute average reversal
rate. We find that an important statistical property of the vortex
reversal, that is, the reversal first starts from peripheral
particles and then gradually to inner particles, so that almost all
particles change their rational directions. Furthermore, we show
that the reversal rate decreases exponentially with the inverse of
noise intensity. Interestingly, the rate varies nonmonotonically
with the number of particles and a minimal rate exists.

We consider $N$ identical SPP in two-dimensional space with
positions $\vec x_i$, velocities $\vec v_i$ ($i=1,\cdots,N$) and
unit mass. The equations of motion read \cite{PhysRevLett.96.104302}
\begin{eqnarray}
&\dot \vec x_i  = \vec v_i, \label{eq1} \\
&\dot \vec v_i  = \left( {\alpha  - \beta \left| {\vec v_i }
\right|^2 } \right)\vec v_i  - \sum\limits_{j \ne i} {\nabla _i
U\left( {\left| {\vec x_i  - \vec x_j } \right|} \right) + \sqrt
{2D} \vec \xi _i }, \label{eq2}
\end{eqnarray}
where the first and second terms on right hand side (rhs) of
Eq.\ref{eq2} represent self-propelled force and friction force,
respectively, and the third term is pair interaction among SPP given
by the generalized Morse potential
\begin{eqnarray}
U\left( {\left| {\vec x_i  - \vec x_j } \right|} \right) = C_r e^{ -
{{\left| {\vec x_i  - \vec x_j } \right|} \mathord{\left/
 {\vphantom {{\left| {\vec x_i  - \vec x_j } \right|} {l_r }}} \right.
 \kern-\nulldelimiterspace} {l_r }}}  - C_a e^{ - {{\left| {\vec x_i  - \vec x_j } \right|} \mathord{\left/
 {\vphantom {{\left| {\vec x_i  - \vec x_j } \right|} {l_a }}} \right.
 \kern-\nulldelimiterspace} {l_a }}}. \label{eq3}
\end{eqnarray}
Here, $l_a$ and $l_r$ represent the attractive and repulsive
potential ranges, $C_a$ and $C_r$ represent their respective
amplitudes. The last term on rhs of Eq.\ref{eq2} is a stochastic
force of intensity $D$ that are independent of particle index and
satisfy $\left\langle { \xi _{i, m} (t)} \right\rangle=0$ and $
\left\langle { \xi _{i,m} (t) \xi _{j,n} (t')} \right\rangle =
\delta _{ij} \delta_{mn} \delta \left( {t - t'} \right)$ with $ i,j
\in 1, \cdots ,N;{\kern 1pt} {\kern 1pt} {\kern 1pt} {\kern 1pt} m,n
\in x,y.$

\begin{figure}
\centerline{\includegraphics*[width=0.8\columnwidth]{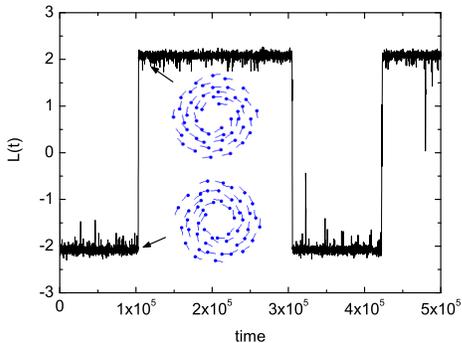}}
\caption{ (Color online) A long-time evolution of $\vec L(t)$ at
$N=40$ and $D=0.37$ shows several vortex reversals. The inset
depicts two typical stable configurations of SPP: vortex with CW
rotation ($\vec L<0$) and CCW rotation ($\vec L>0$).\label{fig1}}
\end{figure}

The model exhibits diverse dynamic patterns, such as clumps, rings
and vortex, for different model parameters. Here we set
$\alpha=1.0$, $\beta=0.5$, $l_a=2.0$, $l_r=0.5$, $C_a=0.6$ and
$C_r=1.0$, which corresponds to the case of vortex. To distinguish
two different rotational directions of vortex, we define average
angular momentum of particles $ \vec L(t)$ as $ \vec L(t)  =
\frac{1} {N}\sum\nolimits_{i = 1}^N {\vec L_i(t)}$, where $\vec
L_i(t)=\vec x_i (t)\times \vec v_i(t) $ is angular momentum of
particle $i$ at time $t$. $ \vec L<0$ and $\vec L>0$ indicate that
SPP rotate clockwise (CW) and counterclockwise (CCW), respectively.
In the presence of a weak noise, vortex pattern is robust to noise.

Figure \ref{fig1} shows a long-time evolution of $\vec L(t)$ at
$N=40$ and $D=0.37$, obtained by numerically integrating
Eqs.\ref{eq1},\ref{eq2} using a fourth order Adams-Bashforth method
\cite{Golub992}, allowing particles an infinite range of motion. One
can observe that noise can induce the sudden transitions of SPP
between two different rational directions. The inset in
Fig.\ref{fig1} depicts two typical configurations of SPP: vortex
with CW and CCW rotation. However, these transitions occur rarely
and average waiting time between transitions is very long. In this
situation, conventional brute-force simulation becomes highly
inefficient. To overcome this difficulty, we will use FFS method of
Allen and coworkers \cite{PhysRevLett.94.018104} to compute the rate
of vortex reversal and evaluate statistical properties of the
reversal path, which is the main purpose of the present work. FFS
method was designed to study rare events both in and out of
equilibrium. This method first defines an order parameter to
distinguish between the initial state $A$ and the final state $B$,
and then uses a series of interfaces to force the system from $A$ to
$B$ in a ratchet-like manner. Here, it is convenient to select $\vec
L$ as the order parameter, and consider CW vortex as $A$ and CCW
vortex as $B$ without loss of generality because these two states
are equivalent in our model.

\begin{figure}
\centerline{\includegraphics*[width=1.0\columnwidth]{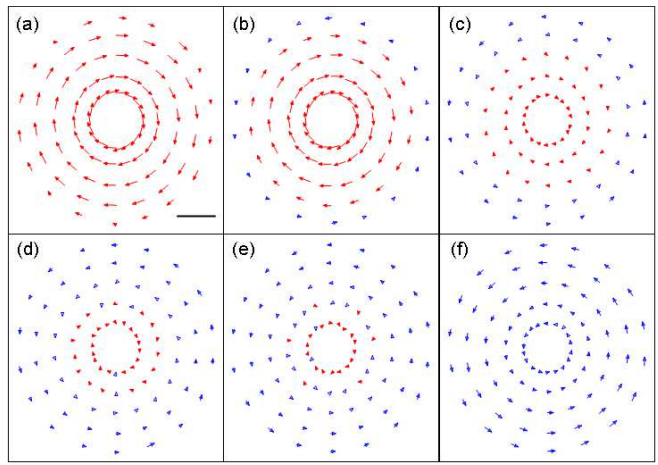}}
\caption{ (Color online) The velocity field of SPP at six different
FFS interfaces, corresponding to (a) $\vec L=-1.95$, (b) $\vec
L=-1.6$, (c) $\vec L=0$, (d) $\vec L=0.25$, (e) $\vec L=0.4$, and
$\vec L=1.0$, respectively. The bar in (a) indicates unit length.
\label{fig2}}
\end{figure}

By storing SPP configurations at each interface of FFS sampling, one
can identify the statistical properties of the vortex reversal
pathway. All results below are obtained by averaging 10 independent
FFS samplings. In each FFS sampling, $1000$ configurations in the
center of mass coordinates are stored at each interface and are
analyzed to obtain the statistical properties of the configurations.
In Fig.\ref{fig2}, we show the velocity field of SPP at six
different interfaces. Fig.\ref{fig2}(a) and Fig.\ref{fig2}(f) show
the velocity field before and after the reversal, indicating that
the system is in stable vortex with CW and CCW rotation,
respectively. While Figs.\ref{fig2}(b-e) show the intermediate
processes of the reversal. One can clearly observe that along the
pathway of vortex reversal the velocity field gradually changes its
sign from the periphery to center of the vortex. That is to say,
vortex reversal first starts from peripheral particles and then
gradually to inner particles, and finally almost all particles
change original rational directions.

\begin{figure}
\centerline{\includegraphics*[width=0.8\columnwidth]{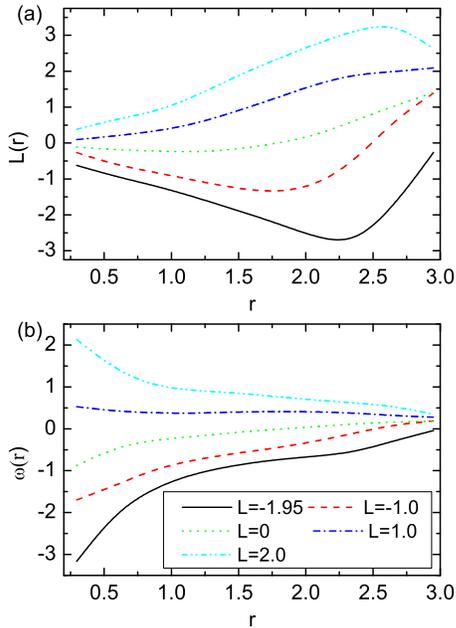}}
\caption{ (Color online) Radial distributions of average angular
momentum $\vec L(r)$ and average angular velocity $\omega (r)$, for
five different interfaces along the pathway of vortex reversal.
\label{fig3}}
\end{figure}

Further information on pathway of vortex reversal is provided by the
radial distributions of average angular momentum $\vec L(r)$ and
average angular velocity $\omega (r)$, where $r$ is the distance to
the center of mass. In Fig.\ref{fig3}, we plot $\vec L(r)$ and
$\omega (r)$ for five different interfaces along the pathway of
vortex reversal. From the variations of $\vec L(r)$ and $\omega (r)$
with interfaces one can observe that the whole process of the vortex
reversal. Before the reversal ($\vec L=-1.95$) the values of $\vec
L(r)$ and $\omega (r)$ are always negative, irrespectively of $r$.
That is, the system is in stable vortex with CW rotation before the
reversal. When the reversal happens, for example, for $\vec L=0$ the
values of $\vec L(r)$ and $\omega (r)$ become positive for $r>1.8$,
while for $r<1.8$ they are always negative. With increasing $\vec L$
this situation further goes till any values of $\vec L(r)$ and
$\omega (r)$ become positive. Finally, the vortex rotates with CCW
direction. Therefore, this further validate that the statistical
property of vortex reversal is that the reversal process starts from
the periphery of the vortex. Also, we calculate the radial
distributions of density of particles and the absolute value of
velocity of particles, and find that they do not have significant
difference with the interfaces $\vec L$.

\begin{figure}
\centerline{\includegraphics*[width=0.8\columnwidth]{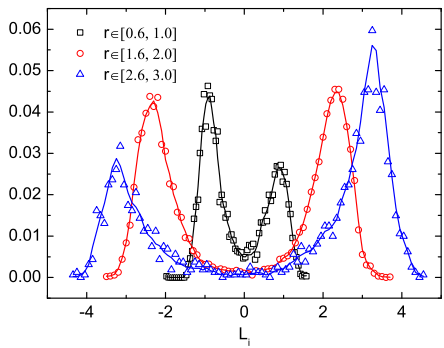}}
\caption{(Color online) The distributions of angular momentum $\vec
L_i$ near the critical configurations for these particles with $r
\in [0.6,1.0]$ (squares), $r \in [1.6 ,2.0]$ (circles), and $r \in
[2.6, 3.0]$ (triangles). \label{fig4}}
\end{figure}

It is informative to analyze the statistical properties of critical
configurations of SPP. Similar to definition in previous studies
\cite{J.Phys.Chem.B10819681}, the critical nucleus is determined by
the committor probability $P_B(i)=0.5$, where $P_B(i)$ is the
probability of reaching $B$ state before returning to $A$ state
starting from the FFS interface $i$, which can be computed by FFS
sampling. This shows that when the system initially locates at the
critical configurations there is equal probability of returning to
CW vortex or CCW vortex. We find that $\vec L\simeq 0.08$ at the
critical configurations, that is, the average angular momentum of
all particles is slightly larger than zero at the critical
configurations. By varying noise intensity $D$ and the number of
particles $N$, the value is not almost changed. Furthermore, we
analyze the radial distributions of angular momentum $\vec L_i$ of
particles near the critical configurations. In Fig.4, we show that
the distributions of $\vec L_i$ of particles at $r \in [0.6,1.0]$
(squares), $[1.6,2.0]$ (circles), and $[2.6,3.0]$ (triangles).
Interestingly, all these distributions are bimodal that are
independent of $r$. The two peaks always correspond to $\vec L_i<0$
and $\vec L_i>0$, respectively, separated by the lowest value of
distributions located at about $\vec L_i\simeq0$. But the loci of
two peaks and their relative heights vary with $r$. For particles at
the inner layer of vortex, the absolute values of $\vec L_i$ is
relatively low, meaning that the rotational property of these
particles is weaken and thus become more disordered. While the other
particles are of obvious ration since the values of $\vec L_i$ are
comparable with ones at stable vortex. The first peak for $\vec
L_i<0$ is higher at the inner layer of vortex, while the second peak
for $\vec L_i<0$ is higher at the outer layer of vortex. At the
center layer of vortex, the heights of the two peaks are nearly the
same. The results give the properties of critical configurations and
thereby further illustrate the mechanism of vortex reversal.

\begin{figure}
\centerline{\includegraphics*[width=1.0\columnwidth]{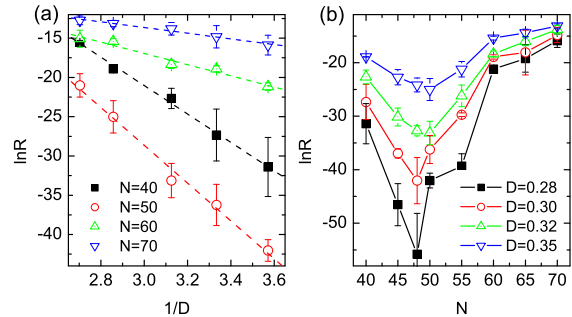}}
\caption{(Color online) The natural logarithm of rate of vortex
reversal $\ln R$ as a function of the inverse of noise intensity
$1/D$ for different SPP number $N$ (a) and as a function of $N$ for
different $D$ (b). The dashed lines in (a) are plotted by linear
fitting. \label{fig5}}
\end{figure}

Another key question in vortex reversal concerns the dependence of
average rate $R$ of the reversal on noise intensity $D$ and the
number of particles $N$. The FFS method can give $R$ as the
production of the flux across the first interface and the
conditional probabilities of reaching the last interface without
returning the first interface. In Fig.\ref{fig5}(a) we plot the
natural logarithm of the rate $\ln R$ as a function of the inverse
of $D$ for $N$. The results show that $R$ decreases exponentially
with $1/D$. It should be pointed out that in our simulation $D$ is
neither too large nor too small. We find that if $D$ is larger than
a critical value $D_c \simeq 0.44$, vortex pattern will be
destroyed, and thus vertex reversal makes no sense. If $D$ is too
small, $R$ is very low so that simulation will be time-consuming. In
Fig.\ref{fig5}(b) we show that $\ln R$ as a function of $N$ for
different $D$. Interestingly, we find that $R$ varies
nonmonotonically with $N$. There exists a minimal value of $R$ at
$N=48$ for low $D$ and at $N=50$ for high $D$. From
Fig.\ref{fig5}(a), one may speculate that there seems to be an
effective nonequilibrium potential for describing the collective
transition in rotational directions. The potential barrier between
CW vortex and CCW vortex is fixed if $N$ is unchanged, such that the
transition rate follow classical Kramers' law. On the other hand,
the nonmonotonic dependence on $R\sim N$ implies dependence of the
effective nonequilibrium potential on $N$ is nontrivial if exists.
However, understanding these results from the view of theoretical
analysis, if not infeasible, is at least a complex task at present.

In summary, using a two-dimensional model of SPP interacting via a
soft-core potential, we have investigated the mechanism of
noise-induced the changes of vortex pattern in rational direction.
By virtue of FFS method we analyze the statistical property and
compute the rate of the reversal. We find that the reversal process
is hierarchical: the process initially inspired by the peripheral
particles, and those particles gradually drive more inner layers of
particles into reverse motion directions. On the other hand, we show
that the rate of the reversal decreases exponentially with the
inverse of noise intensity. Interestingly, the reversal rate depends
nonmonotonically on the number of SPP and a minimal rate exists at a
moderate number of particles. Our findings may provide us some new
understanding on the transitions of collective patterns of SPP.

\begin{acknowledgments}
This work is supported by NSFC (Grant Nos. 91027012, 20933006). Z.H.
acknowledges support by China National Funds for Distinguished Young
Scientists (Grant No. 21125313). H.C. acknowledges support by the
Doctoral Research Foundation of Anhui University (Grant No.
KJ2012B042).

\end{acknowledgments}

%

\begin{thebibliography}{30}
\expandafter\ifx\csname
natexlab\endcsname\relax\def\natexlab#1{#1}\fi
\expandafter\ifx\csname bibnamefont\endcsname\relax
  \def\bibnamefont#1{#1}\fi
\expandafter\ifx\csname bibfnamefont\endcsname\relax
  \def\bibfnamefont#1{#1}\fi
\expandafter\ifx\csname citenamefont\endcsname\relax
  \def\citenamefont#1{#1}\fi
\expandafter\ifx\csname url\endcsname\relax
  \def\url#1{\texttt{#1}}\fi
\expandafter\ifx\csname urlprefix\endcsname\relax\def\urlprefix{URL
}\fi \providecommand{\bibinfo}[2]{#2}
\providecommand{\eprint}[2][]{\url{#2}}

\bibitem[{\citenamefont{Vicsek and Zafiris}(2010)}]{arXiv1010.5017}
\bibinfo{author}{\bibfnamefont{T.}~\bibnamefont{Vicsek}} \bibnamefont{and}
  \bibinfo{author}{\bibfnamefont{A.}~\bibnamefont{Zafiris}},
  \bibinfo{journal}{e-print: arXiv:1010.5017}  (\bibinfo{year}{2010}).

\bibitem[{\citenamefont{Helbing}(2001)}]{RevModPhys.73.1067}
\bibinfo{author}{\bibfnamefont{D.}~\bibnamefont{Helbing}},
  \bibinfo{journal}{Rev. Mod. Phys.} \textbf{\bibinfo{volume}{73}},
  \bibinfo{pages}{1067} (\bibinfo{year}{2001}).

\bibitem[{\citenamefont{\emph{et al.}}(2010)}]{PNAS107.11865}
\bibinfo{author}{\bibfnamefont{A.~C.} \bibnamefont{\emph{et al.}}},
  \bibinfo{journal}{Proc. Natl. Acad. Sci. U.S.A.}
  \textbf{\bibinfo{volume}{107}}, \bibinfo{pages}{11865}
  (\bibinfo{year}{2010}).

\bibitem[{\citenamefont{Bhattacharya and Vicsek}(2010)}]{NJP12.093019}
\bibinfo{author}{\bibfnamefont{K.}~\bibnamefont{Bhattacharya}}
  \bibnamefont{and} \bibinfo{author}{\bibfnamefont{T.}~\bibnamefont{Vicsek}},
  \bibinfo{journal}{New J. Phys.} \textbf{\bibinfo{volume}{12}},
  \bibinfo{pages}{093019} (\bibinfo{year}{2010}).

\bibitem[{\citenamefont{\emph{et al.}}(2006)}]{Science312.1402}
\bibinfo{author}{\bibfnamefont{J.~B.} \bibnamefont{\emph{et al.}}},
  \bibinfo{journal}{Science} \textbf{\bibinfo{volume}{312}},
  \bibinfo{pages}{1402} (\bibinfo{year}{2006}).

\bibitem[{\citenamefont{Romanczuk et~al.}(2009)\citenamefont{Romanczuk, Couzin,
  and Schimansky-Geier}}]{PhysRevLett.102.010602}
\bibinfo{author}{\bibfnamefont{P.}~\bibnamefont{Romanczuk}},
  \bibinfo{author}{\bibfnamefont{I.~D.} \bibnamefont{Couzin}},
  \bibnamefont{and}
  \bibinfo{author}{\bibfnamefont{L.}~\bibnamefont{Schimansky-Geier}},
  \bibinfo{journal}{Phys. Rev. Lett.} \textbf{\bibinfo{volume}{102}},
  \bibinfo{pages}{010602} (\bibinfo{year}{2009}).

\bibitem[{\citenamefont{Angelani et~al.}(2009)\citenamefont{Angelani,
  Di~Leonardo, and Ruocco}}]{PhysRevLett.102.048104}
\bibinfo{author}{\bibfnamefont{L.}~\bibnamefont{Angelani}},
  \bibinfo{author}{\bibfnamefont{R.}~\bibnamefont{Di~Leonardo}},
  \bibnamefont{and} \bibinfo{author}{\bibfnamefont{G.}~\bibnamefont{Ruocco}},
  \bibinfo{journal}{Phys. Rev. Lett.} \textbf{\bibinfo{volume}{102}},
  \bibinfo{pages}{048104} (\bibinfo{year}{2009}).

\bibitem[{\citenamefont{H.~P.~Zhang and Swinney}(2010)}]{PNAS107.13626}
\bibinfo{author}{\bibfnamefont{E.-L.~F.} \bibnamefont{H.~P.~Zhang},
  \bibfnamefont{Avraham~Be'er}} \bibnamefont{and}
  \bibinfo{author}{\bibfnamefont{H.~L.} \bibnamefont{Swinney}},
  \bibinfo{journal}{Proc. Natl. Acad. Sci. U.S.A.}
  \textbf{\bibinfo{volume}{107}}, \bibinfo{pages}{13626}
  (\bibinfo{year}{2010}).

\bibitem[{\citenamefont{V.~Narayan and Menon}(2007)}]{Science317.105}
\bibinfo{author}{\bibfnamefont{S.~R.} \bibnamefont{V.~Narayan}}
  \bibnamefont{and} \bibinfo{author}{\bibfnamefont{N.}~\bibnamefont{Menon}},
  \bibinfo{journal}{Science} \textbf{\bibinfo{volume}{317}},
  \bibinfo{pages}{105} (\bibinfo{year}{2007}).

\bibitem[{\citenamefont{Deseigne et~al.}(2010)\citenamefont{Deseigne, Dauchot,
  and Chat\'e}}]{PhysRevLett.105.098001}
\bibinfo{author}{\bibfnamefont{J.}~\bibnamefont{Deseigne}},
  \bibinfo{author}{\bibfnamefont{O.}~\bibnamefont{Dauchot}}, \bibnamefont{and}
  \bibinfo{author}{\bibfnamefont{H.}~\bibnamefont{Chat\'e}},
  \bibinfo{journal}{Phys. Rev. Lett.} \textbf{\bibinfo{volume}{105}},
  \bibinfo{pages}{098001} (\bibinfo{year}{2010}).

\bibitem[{\citenamefont{Vicsek et~al.}(1995)\citenamefont{Vicsek, Czir\'ok,
  Ben-Jacob, Cohen, and Shochet}}]{PhysRevLett.75.1226}
\bibinfo{author}{\bibfnamefont{T.}~\bibnamefont{Vicsek}},
  \bibinfo{author}{\bibfnamefont{A.}~\bibnamefont{Czir\'ok}},
  \bibinfo{author}{\bibfnamefont{E.}~\bibnamefont{Ben-Jacob}},
  \bibinfo{author}{\bibfnamefont{I.}~\bibnamefont{Cohen}}, \bibnamefont{and}
  \bibinfo{author}{\bibfnamefont{O.}~\bibnamefont{Shochet}},
  \bibinfo{journal}{Phys. Rev. Lett.} \textbf{\bibinfo{volume}{75}},
  \bibinfo{pages}{1226} (\bibinfo{year}{1995}).

\bibitem[{\citenamefont{Toner and Tu}(1995)}]{PhysRevLett.75.4326}
\bibinfo{author}{\bibfnamefont{J.}~\bibnamefont{Toner}} \bibnamefont{and}
  \bibinfo{author}{\bibfnamefont{Y.}~\bibnamefont{Tu}}, \bibinfo{journal}{Phys.
  Rev. Lett.} \textbf{\bibinfo{volume}{75}}, \bibinfo{pages}{4326}
  (\bibinfo{year}{1995}).

\bibitem[{\citenamefont{Gr\'egoire and Chat\'e}(2004)}]{PhysRevLett.92.025702}
\bibinfo{author}{\bibfnamefont{G.}~\bibnamefont{Gr\'egoire}} \bibnamefont{and}
  \bibinfo{author}{\bibfnamefont{H.}~\bibnamefont{Chat\'e}},
  \bibinfo{journal}{Phys. Rev. Lett.} \textbf{\bibinfo{volume}{92}},
  \bibinfo{pages}{025702} (\bibinfo{year}{2004}).

\bibitem[{\citenamefont{Aldana et~al.}(2007)\citenamefont{Aldana, Dossetti,
  Huepe, Kenkre, and Larralde}}]{PhysRevLett.98.095702}
\bibinfo{author}{\bibfnamefont{M.}~\bibnamefont{Aldana}},
  \bibinfo{author}{\bibfnamefont{V.}~\bibnamefont{Dossetti}},
  \bibinfo{author}{\bibfnamefont{C.}~\bibnamefont{Huepe}},
  \bibinfo{author}{\bibfnamefont{V.~M.} \bibnamefont{Kenkre}},
  \bibnamefont{and} \bibinfo{author}{\bibfnamefont{H.}~\bibnamefont{Larralde}},
  \bibinfo{journal}{Phys. Rev. Lett.} \textbf{\bibinfo{volume}{98}},
  \bibinfo{pages}{095702} (\bibinfo{year}{2007}).

\bibitem[{\citenamefont{Chat\'e et~al.}(2006)\citenamefont{Chat\'e, Ginelli,
  and Montagne}}]{PhysRevLett.96.180602}
\bibinfo{author}{\bibfnamefont{H.}~\bibnamefont{Chat\'e}},
  \bibinfo{author}{\bibfnamefont{F.}~\bibnamefont{Ginelli}}, \bibnamefont{and}
  \bibinfo{author}{\bibfnamefont{R.}~\bibnamefont{Montagne}},
  \bibinfo{journal}{Phys. Rev. Lett.} \textbf{\bibinfo{volume}{96}},
  \bibinfo{pages}{180602} (\bibinfo{year}{2006}).

\bibitem[{\citenamefont{Levine et~al.}(2000)\citenamefont{Levine, Rappel, and
  Cohen}}]{PhysRevE.63.017101}
\bibinfo{author}{\bibfnamefont{H.}~\bibnamefont{Levine}},
  \bibinfo{author}{\bibfnamefont{W.-J.} \bibnamefont{Rappel}},
  \bibnamefont{and} \bibinfo{author}{\bibfnamefont{I.}~\bibnamefont{Cohen}},
  \bibinfo{journal}{Phys. Rev. E} \textbf{\bibinfo{volume}{63}},
  \bibinfo{pages}{017101} (\bibinfo{year}{2000}).

\bibitem[{\citenamefont{D'Orsogna et~al.}(2006)\citenamefont{D'Orsogna, Chuang,
  Bertozzi, and Chayes}}]{PhysRevLett.96.104302}
\bibinfo{author}{\bibfnamefont{M.~R.} \bibnamefont{D'Orsogna}},
  \bibinfo{author}{\bibfnamefont{Y.~L.} \bibnamefont{Chuang}},
  \bibinfo{author}{\bibfnamefont{A.~L.} \bibnamefont{Bertozzi}},
  \bibnamefont{and} \bibinfo{author}{\bibfnamefont{L.~S.}
  \bibnamefont{Chayes}}, \bibinfo{journal}{Phys. Rev. Lett.}
  \textbf{\bibinfo{volume}{96}}, \bibinfo{pages}{104302}
  (\bibinfo{year}{2006}).

\bibitem[{\citenamefont{Peruani et~al.}(2011)\citenamefont{Peruani, Klauss,
  Deutsch, and Voss-Boehme}}]{PhysRevLett.106.128101}
\bibinfo{author}{\bibfnamefont{F.}~\bibnamefont{Peruani}},
  \bibinfo{author}{\bibfnamefont{T.}~\bibnamefont{Klauss}},
  \bibinfo{author}{\bibfnamefont{A.}~\bibnamefont{Deutsch}}, \bibnamefont{and}
  \bibinfo{author}{\bibfnamefont{A.}~\bibnamefont{Voss-Boehme}},
  \bibinfo{journal}{Phys. Rev. Lett.} \textbf{\bibinfo{volume}{106}},
  \bibinfo{pages}{128101} (\bibinfo{year}{2011}).

\bibitem[{\citenamefont{Yatesa et~al.}(2009)\citenamefont{Yatesa, Erbana,
  Escudero, Couzin, Buhl, Kevrekidis, Mainia, and T}}]{PNAS106.5464}
\bibinfo{author}{\bibfnamefont{C.~A.} \bibnamefont{Yatesa}},
  \bibinfo{author}{\bibfnamefont{R.}~\bibnamefont{Erbana}},
  \bibinfo{author}{\bibfnamefont{C.}~\bibnamefont{Escudero}},
  \bibinfo{author}{\bibfnamefont{I.~D.} \bibnamefont{Couzin}},
  \bibinfo{author}{\bibfnamefont{J.}~\bibnamefont{Buhl}},
  \bibinfo{author}{\bibfnamefont{I.~G.} \bibnamefont{Kevrekidis}},
  \bibinfo{author}{\bibfnamefont{P.~K.} \bibnamefont{Mainia}},
  \bibnamefont{and} \bibinfo{author}{\bibfnamefont{S.~D.~J.} \bibnamefont{T}},
  \bibinfo{journal}{Proc. Natl. Acad. Sci. U.S.A.}
  \textbf{\bibinfo{volume}{106}}, \bibinfo{pages}{5464} (\bibinfo{year}{2009}).

\bibitem[{\citenamefont{Mikhailov and Zanette}(1999)}]{PhysRevE.60.4571}
\bibinfo{author}{\bibfnamefont{A.~S.} \bibnamefont{Mikhailov}}
  \bibnamefont{and} \bibinfo{author}{\bibfnamefont{D.~H.}
  \bibnamefont{Zanette}}, \bibinfo{journal}{Phys. Rev. E}
  \textbf{\bibinfo{volume}{60}}, \bibinfo{pages}{4571} (\bibinfo{year}{1999}).

\bibitem[{\citenamefont{Erdmann et~al.}(2005)\citenamefont{Erdmann, Ebeling,
  and Mikhailov}}]{PhysRevE.71.051904}
\bibinfo{author}{\bibfnamefont{U.}~\bibnamefont{Erdmann}},
  \bibinfo{author}{\bibfnamefont{W.}~\bibnamefont{Ebeling}}, \bibnamefont{and}
  \bibinfo{author}{\bibfnamefont{A.~S.} \bibnamefont{Mikhailov}},
  \bibinfo{journal}{Phys. Rev. E} \textbf{\bibinfo{volume}{71}},
  \bibinfo{pages}{051904} (\bibinfo{year}{2005}).

\bibitem[{\citenamefont{Allison~Kolpas and Kevrekidis}(2007)}]{PNAS104.5931}
\bibinfo{author}{\bibfnamefont{J.~M.} \bibnamefont{Allison~Kolpas}}
  \bibnamefont{and} \bibinfo{author}{\bibfnamefont{I.~G.}
  \bibnamefont{Kevrekidis}}, \bibinfo{journal}{Proc. Natl. Acad. Sci. U.S.A.}
  \textbf{\bibinfo{volume}{104}}, \bibinfo{pages}{5931} (\bibinfo{year}{2007}).

\bibitem[{\citenamefont{Strefler et~al.}(2008)\citenamefont{Strefler, Erdmann,
  and Schimansky-Geier}}]{PhysRevE.78.031927}
\bibinfo{author}{\bibfnamefont{J.}~\bibnamefont{Strefler}},
  \bibinfo{author}{\bibfnamefont{U.}~\bibnamefont{Erdmann}}, \bibnamefont{and}
  \bibinfo{author}{\bibfnamefont{L.}~\bibnamefont{Schimansky-Geier}},
  \bibinfo{journal}{Phys. Rev. E} \textbf{\bibinfo{volume}{78}},
  \bibinfo{pages}{031927} (\bibinfo{year}{2008}).

\bibitem[{\citenamefont{Julia K.~Parrish and Grunbaum}(2002)}]{BiolBull202.296}
\bibinfo{author}{\bibfnamefont{S.~V.~V.} \bibnamefont{Julia K.~Parrish}}
  \bibnamefont{and} \bibinfo{author}{\bibfnamefont{D.}~\bibnamefont{Grunbaum}},
  \bibinfo{journal}{Biol. Bull.} \textbf{\bibinfo{volume}{202}},
  \bibinfo{pages}{296} (\bibinfo{year}{2002}).

\bibitem[{\citenamefont{Sumpter}(2006)}]{PhilTransRSocB361.5}
\bibinfo{author}{\bibfnamefont{D.~J.~T.} \bibnamefont{Sumpter}},
  \bibinfo{journal}{Phil. Trans. R. Soc. B} \textbf{\bibinfo{volume}{361}},
  \bibinfo{pages}{5} (\bibinfo{year}{2006}).

\bibitem[{\citenamefont{Czir\'ok et~al.}(1996)\citenamefont{Czir\'ok,
  Ben-Jacob, Cohen, and Vicsek}}]{PhysRevE.54.1791}
\bibinfo{author}{\bibfnamefont{A.}~\bibnamefont{Czir\'ok}},
  \bibinfo{author}{\bibfnamefont{E.}~\bibnamefont{Ben-Jacob}},
  \bibinfo{author}{\bibfnamefont{I.}~\bibnamefont{Cohen}}, \bibnamefont{and}
  \bibinfo{author}{\bibfnamefont{T.}~\bibnamefont{Vicsek}},
  \bibinfo{journal}{Phys. Rev. E} \textbf{\bibinfo{volume}{54}},
  \bibinfo{pages}{1791} (\bibinfo{year}{1996}).

\bibitem[{\citenamefont{Rappel et~al.}(1999)\citenamefont{Rappel, Nicol,
  Sarkissian, Levine, and Loomis}}]{PhysRevLett.83.1247}
\bibinfo{author}{\bibfnamefont{W.-J.} \bibnamefont{Rappel}},
  \bibinfo{author}{\bibfnamefont{A.}~\bibnamefont{Nicol}},
  \bibinfo{author}{\bibfnamefont{A.}~\bibnamefont{Sarkissian}},
  \bibinfo{author}{\bibfnamefont{H.}~\bibnamefont{Levine}}, \bibnamefont{and}
  \bibinfo{author}{\bibfnamefont{W.~F.} \bibnamefont{Loomis}},
  \bibinfo{journal}{Phys. Rev. Lett.} \textbf{\bibinfo{volume}{83}},
  \bibinfo{pages}{1247} (\bibinfo{year}{1999}).

\bibitem[{\citenamefont{Allen et~al.}(2005)\citenamefont{Allen, Warren, and ten
  Wolde}}]{PhysRevLett.94.018104}
\bibinfo{author}{\bibfnamefont{R.~J.} \bibnamefont{Allen}},
  \bibinfo{author}{\bibfnamefont{P.~B.} \bibnamefont{Warren}},
  \bibnamefont{and} \bibinfo{author}{\bibfnamefont{P.~R.} \bibnamefont{ten
  Wolde}}, \bibinfo{journal}{Phys. Rev. Lett.} \textbf{\bibinfo{volume}{94}},
  \bibinfo{pages}{018104} (\bibinfo{year}{2005}).

\bibitem[{\citenamefont{Golub and Ortega}(1992)}]{Golub992}
\bibinfo{author}{\bibfnamefont{G.~H.} \bibnamefont{Golub}} \bibnamefont{and}
  \bibinfo{author}{\bibfnamefont{J.~M.} \bibnamefont{Ortega}},
  \emph{\bibinfo{title}{Scientific Computing and Differential Equations: An
  Introduction to Numerical Methods}} (\bibinfo{publisher}{Academic Press},
  \bibinfo{year}{1992}).

\bibitem[{\citenamefont{Pan and Chandler}(2004)}]{J.Phys.Chem.B10819681}
\bibinfo{author}{\bibfnamefont{A.~C.} \bibnamefont{Pan}} \bibnamefont{and}
  \bibinfo{author}{\bibfnamefont{D.}~\bibnamefont{Chandler}},
  \bibinfo{journal}{J. Phys. Chem. B} \textbf{\bibinfo{volume}{108}},
  \bibinfo{pages}{19681} (\bibinfo{year}{2004}).

\end{thebibliography}

\end{document}